\documentclass[aps,prl,reprint,twocolumn,showpacs,showkeys]{revtex4-1}

\usepackage{graphicx}

\newcommand{\vc}[1]{\textbf{\em #1}}

\newcommand{\aap}{Astron. Astrophys.}
\newcommand{\apjl}{Astrophys. J. Lett.}
\newcommand{\app}{Astropart. Phys.}
\newcommand{\araa}{Annu. Rev. Astron. Astrophys.}
\newcommand{\grl}{Geophys. Res. Lett.}
\newcommand{\ijmpd}{Int. J. Mod. Phys.}
\newcommand{\ijnme}{Int. J. Num. Met. Eng.}
\newcommand{\jcph}{J. Comp. Phys.}
\newcommand{\jgr}{J. Geophys. Res.}
\newcommand{\jpcs}{J. Phys. Conf. Ser.}
\newcommand{\mnras}{Mon. Not. R. Astron. Soc.}
\newcommand{\npg}{Nonlin. Proc. Geophys.}
\newcommand{\pop}{Phys. Plas.}
\newcommand{\ssr}{Space Sc. Rev.}

\begin{document}

\title{Particle Acceleration in Turbulence and Weakly Stochastic Reconnection}

\author{Grzegorz Kowal}
\email{kowal@astro.iag.usp.br}
\author{Elisabete M. de Gouveia Dal Pino}
\affiliation{Instituto de Astronomia, Geof\'\i sica e Ci\^encias Atmosf\'ericas,
  Universidade de S\~ao Paulo, Rua do Mat\~ao, 1226 -- Cidade Universit\'{a}ria,
  CEP 05508-090, S\~ao Paulo/SP, Brazil}
\author{Alex Lazarian}
\affiliation{Department of Astronomy, University of Wisconsin,
  475 North Charter Street, Madison, WI 53706, USA}

\begin{abstract}
Fast particles are accelerated in astrophysical environments by a variety of
processes.  Acceleration in reconnection sites has attracted the attention of
researchers recently.  In this letter we analyze the energy distribution
evolution of test particles injected in three dimensional (3D)
magnetohydrodynamic (MHD) simulations of different magnetic reconnection
configurations.  When considering a single Sweet-Parker topology, the particles
accelerate predominantly through a first-order Fermi process, as predicted in
previous work \cite{degouveia05} and demonstrated numerically in \cite{kowal11}.
When turbulence is included within the current sheet, the acceleration rate,
which depends on the reconnection rate, is highly enhanced.  This is because
reconnection in the presence of turbulence becomes fast and independent of
resistivity \cite{lazarian99, kowal09} and allows the formation of a thick
volume filled with multiple simultaneously reconnecting magnetic fluxes. Charged
particles trapped within this volume suffer several head-on scatterings with the
contracting magnetic fluctuations, which significantly increase the acceleration
rate and results in a first-order Fermi process.  For comparison, we also tested
acceleration in MHD turbulence, where particles suffer collisions with
approaching and receding magnetic irregularities, resulting in a reduced
acceleration rate.  We argue that the dominant acceleration mechanism approaches
a second order Fermi process in this case.
\end{abstract}



\maketitle

\paragraph{Introduction.}

Energetic particles are ubiquitous in astrophysical environments.  Cosmic ray
(CR) acceleration still challenges the researchers.  For instance, the origin of
ultra high energy cosmic rays (UHECRs) is still unknown, although several
mechanism have been proposed \cite[][and references there in]{ostrowski02}.
Their spectrum is consistent with an origin in extragalactic astrophysical
sources and candidates range from the birth of compact objects to explosions
related to gamma-ray bursts (GRBs) or events in active galaxies (AGNs)
\cite{kotera11}.  Very high energy observations of AGNs and GRBs with the Fermi
and Swift satellites and ground based gamma ray observatories (HESS, VERITAS and
MAGIC) challenge current theories of particle acceleration, mostly based on the
acceleration in shocks, trying to explain how particles are accelerated to
energies above TeV in regions relatively small compared to the fiducial scale of
their sources \cite{sol11}.

While particle acceleration in shocks has been extensively explored
\cite{sironi09} (see also \cite{melrose09, kotera11} for reviews), an
alternative, less investigated mechanism so far, involves particle acceleration
in magnetic reconnection sites.  Magnetic reconnection may occur when two
magnetic fluxes of opposite polarity encounter each other.  In the presence of
finite magnetic diffusivity the converging magnetic lines annihilate at the
discontinuity surface and a current sheet forms there.  In \cite{degouveia05},
the authors first proposed that an efficient first-order Fermi process can occur
within a current sheet, where trapped charged particles may bounce back and
forth several times and gain energy due to head-on collisions with the two
converging magnetic fluxes incoming with the reconnection speed $V_{rec}$.  They
found that the particle energy gain after each round trip is $\Delta E/E \propto
V_{rec}/c$.  With a {\em fast} magnetic reconnection, like the one induced in
the presence of turbulence \cite{lazarian99}, $V_{rec}$ can be of the order of
Alfv\'en speed $V_{A}$.  Afterwards, \cite{drake06} appealed to a similar
process, but within a collisionless reconnection scenario.  In their model, the
acceleration is controlled by the contraction of two-dimensional (2D) loops due
to firehose instability that arises in a particle-in-cell (PIC) domain \cite[see
also][]{lyubarsky08,drake10}.  Other processes of acceleration, like those due
to the electric field associated with the current in the reconnection region
\cite{litvinenko96} and turbulence arising as a result of reconnection
\cite{larosa06}, were shown to be less dominant.

Magnetic reconnection is expected to induce acceleration of particles in a wide
range of galactic and extragalactic environments.  Originally discussed
predominantly in the context of solar flares \cite[e.g.][]{drake06, drake09,
gordovskyy10, gordovskyy11, zharkova11, lazarian09, drake10, lazarian10}, it has
been recently extended to more extreme astrophysical environments and sources,
such as the production of UHECRs \citep{degouveia00, degouveia01, kotera11}, in
particle acceleration in jet-accretion disk systems \citep{degouveia10a,
degouveia10b, giannios10, delvalle11}, and in the general framework of AGNs and
GRBs \citep{degouveia10b, giannios10, zhang11, uzdensky11a, uzdensky11b}.  These
applications, however, still require in-depth studies of particle acceleration
in magnetic reconnection sites, and its connection with magnetohydrodynamical
(MHD) turbulence and {\em fast} magnetic reconnection.

In this letter, we explore this issue by means of fully 3D MHD simulations of
particle acceleration in magnetic reconnection domains.  A preliminary study in
this direction was performed in \cite{kowal11} where the analytical model of
\cite{degouveia05} was successfully tested.  It was also shown that particle
acceleration taking place in MHD reconnection domains without including kinetic
effects produces results similar to those found in collisionless PIC simulations
\citep{drake10}.  This proved that the acceleration in reconnection regions is a
universal process which is not determined by the details of the plasma physics
and can be very efficient in collisional gas, although energy and radiative
losses due to the interactions of the accelerated particles with the surrounding
plasma may be significant in some systems.  Moreover, it has been shown that
particle acceleration in 2D and 3D MHD reconnection behaves quite differently,
what calls for focusing on realistic 3D geometries.

In this work we study differences between energy distributions of test particles
injected in 3D MHD reconnection sites with and without turbulence, corresponding
to the Sweet-Parker \cite{sweet58,parker57} and Lazarian-Vishniac
\cite{lazarian99} configurations.  For comparison, we also considered a pure
turbulent environment.

\paragraph{Numerical Simulations of Reconnection and Turbulence.}
\label{sec:simulations}

Magnetic reconnection and turbulence were modeled by solving the MHD equations
numerically on a uniform mesh using a shock-capturing Godunov-type scheme based
on the 2$^{nd}$ order spatial reconstruction and 2$^{nd}$ Runge-Kutta (RK) time
integration \cite{kowal09}.  We used the isothermal HLLD Riemann solver
\cite{mignone07} to obtain the numerical fluxes in the update step, what
resulted in reducing the energy dissipation of Alfv\'en waves.  We incorporated
the field-interpolated constrained transport (CT) scheme based on a staggered
mesh \cite{londrillo00} into the integration of the induction equation to
maintain the $\nabla \cdot \vc{B} = 0$ constraint numerically.

In the reconnection models we adopted an initial Harris current sheet,
$B_x(x,y,z) = B_{0x} \tanh (y/\theta)$, initialized by the magnetic vector
potential with a uniform guide field $B_z(x,y,z) = B_{0z} = \mathrm{const}$ and
the density profile set from the uniform total (thermal plus magnetic) pressure
$p_T(x,y,z) = \mathrm{const}$.  Initial velocity was zero everywhere.  The
reconnection was initiated by a small initial perturbation $\delta A_z(x,y,z) =
\delta B_{0x} \cos(2 \pi x) \exp[-(y/d)^2]$ added to the initial vector
potential.  The parameters $\delta B_{0x}$ and $d$ describe the strength of the
initial perturbation and thickness of the perturbed region, respectively.  In
the case of turbulent model the initial magnetic field was uniform.

We employed dimensionless equations, so that the velocity and magnetic field are
expressed in the fiducial Alfv\'en speed units (defined by the antiparallel
component of the  field) and the unperturbed density $\rho_0=1$).  The length of
the box in the X direction defines the distance unit and the time is measured in
units of $L/V_A$.  Initially, we set the strength of the antiparallel magnetic
field component $B_{0x}$ to 1.0 and the guide field $B_{0z}$ to 0.1.  The sound
speed was set to 4.0 in all models in order to suppress the compressibility in
the system.  The parameters describing the initial perturbation are set to
$\delta B_{0x} = 0.05$ and $d = 0.1$.  For more details of the computational
method and numerical setup see \cite{kowal09,kowal11b}.

\paragraph{Integration of Particle Trajectories.}
\label{sec:methods}

In order to integrate the test particle trajectories we freeze in time a data
cube obtained from the MHD models and inject test particles in the domain with
random initial positions and directions and with an initial thermal
distribution.  For each particle we solve the relativistic motion equation
\begin{equation}
 \frac{d}{d t} \left( \gamma m \vc{u} \right) = q \left( \vc{E} + \vc{u} \times
 \vc{B} \right) , \label{eq:ptrajectory}
\end{equation}
where $m$, $q$ and $\vc{u}$ are the particle mass, electric charge and velocity,
respectively, $\vc{E}$ and $\vc{B}$ are the electric and magnetic fields,
respectively, $\gamma \equiv \left( 1 - u^2 / c^2 \right)^{-1}$ is the Lorentz
factor, and $c$ is the speed of light.  The electric field $\vc{E}$ is taken
from the MHD simulations
\begin{equation}
 \vc{E} = - \vc{v} \times \vc{B} + \eta \vc{J} , \label{eq:efield}
\end{equation}
where $\vc{v}$ is the plasma velocity, $\vc{J} \equiv \nabla \times \vc{B}$ is
the current density, and $\eta$ is the Ohmic resistivity coefficient.  We
neglect the resistive term above since its effect on particle acceleration is
negligible \cite{kowal11}.  In the current studies we do not include the
particle energy losses, thus particles can gain or loose energy only through the
interactions with the moving magnetized plasma.  The inclusion of radiative
losses or back reaction on the plasma is planned for future studies,
particularly with the incorporation of the kinetic effects.  We note that since
we are focusing on the acceleration process only, we consider very simple
domains which represent only small periodic boxes of entire magnetic
reconnection or turbulent sites.  For this reason, the typical crossing time
through the box of an injected thermal particle is very small and it has to
re-enter the computational domain several times before gaining significant
energy by multiple scatterings.  Thus, whenever a particle reaches the box
boundary it re-enters in the other side to continue scattering.

Eq.~\ref{eq:ptrajectory} is integrated using the 4$^{th}$ order Runge-Kutta
method with the adaptive time step based on the 5$^{th}$ order error estimator
\cite[see][e.g.]{press92}.  The fields are interpolated using the 2$^{nd}$ order
methods \citep{lekien05} and limited to 0$^{th}$ order near discontinuities
\cite{kowal11}.

For convenience, we assume the speed of light to be 20 $V_A$, which defines our
plasma in a non-relativistic regime, and the mean density is assumed to be 1
atomic mass unit per cubic centimeter, which is a fiducial value, e.g., of the
interstellar medium (ISM) density.  All times are expressed in units of the
Alfv\'en time.

\paragraph{Acceleration in Sweet-Parker Reconnection.}
\label{sec:sweet-parker-reconnection}

The acceleration of a single test particle in the Sweet-Parker reconnection was
already described in \cite{kowal11}.  Here, we present statistical studies for
10,000 protons injected in such a domain.

\begin{figure*}
\center
\includegraphics[width=\columnwidth]{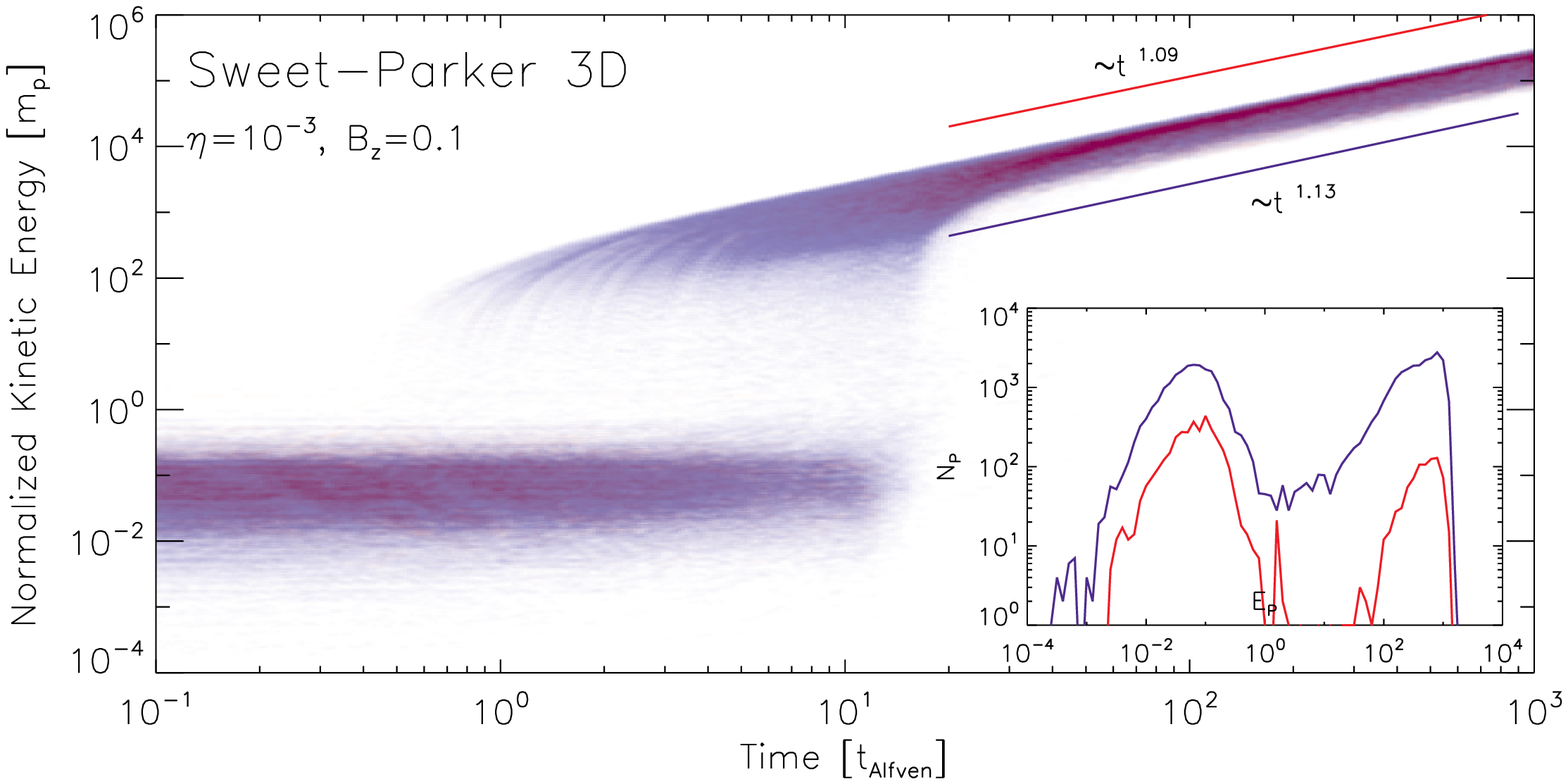}
\includegraphics[width=\columnwidth]{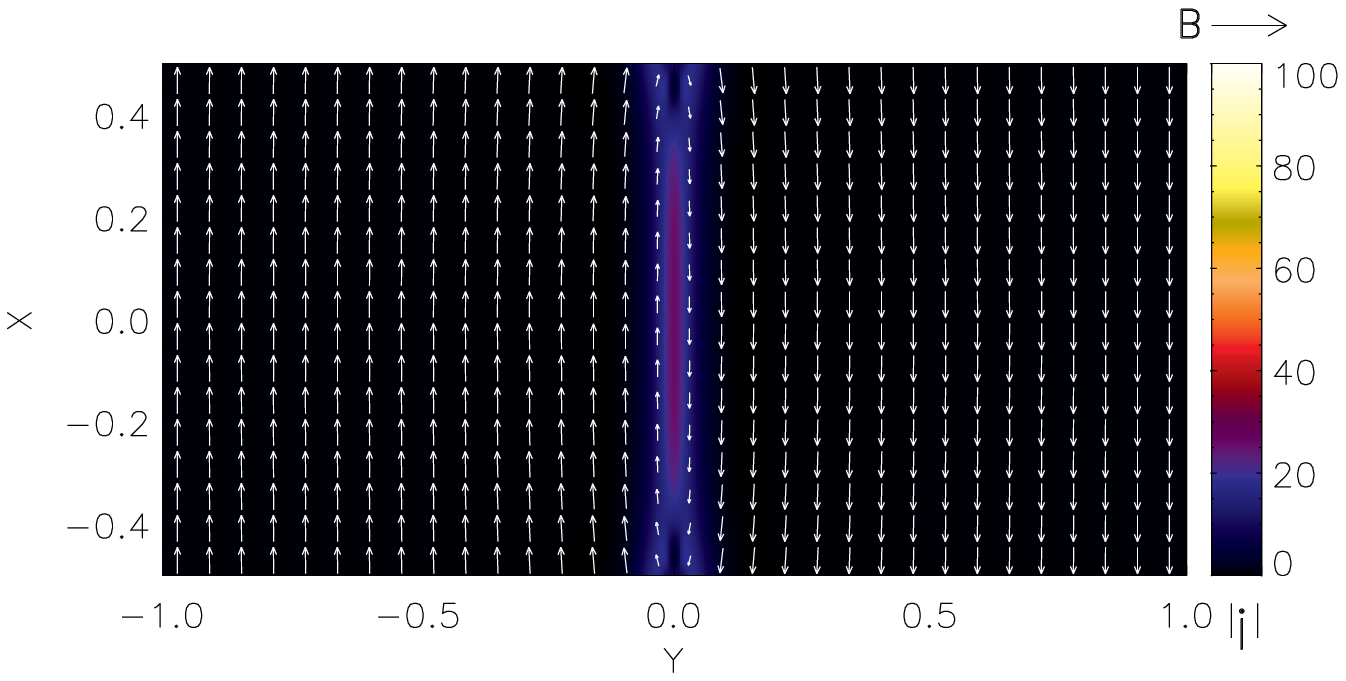} \\
\includegraphics[width=\columnwidth]{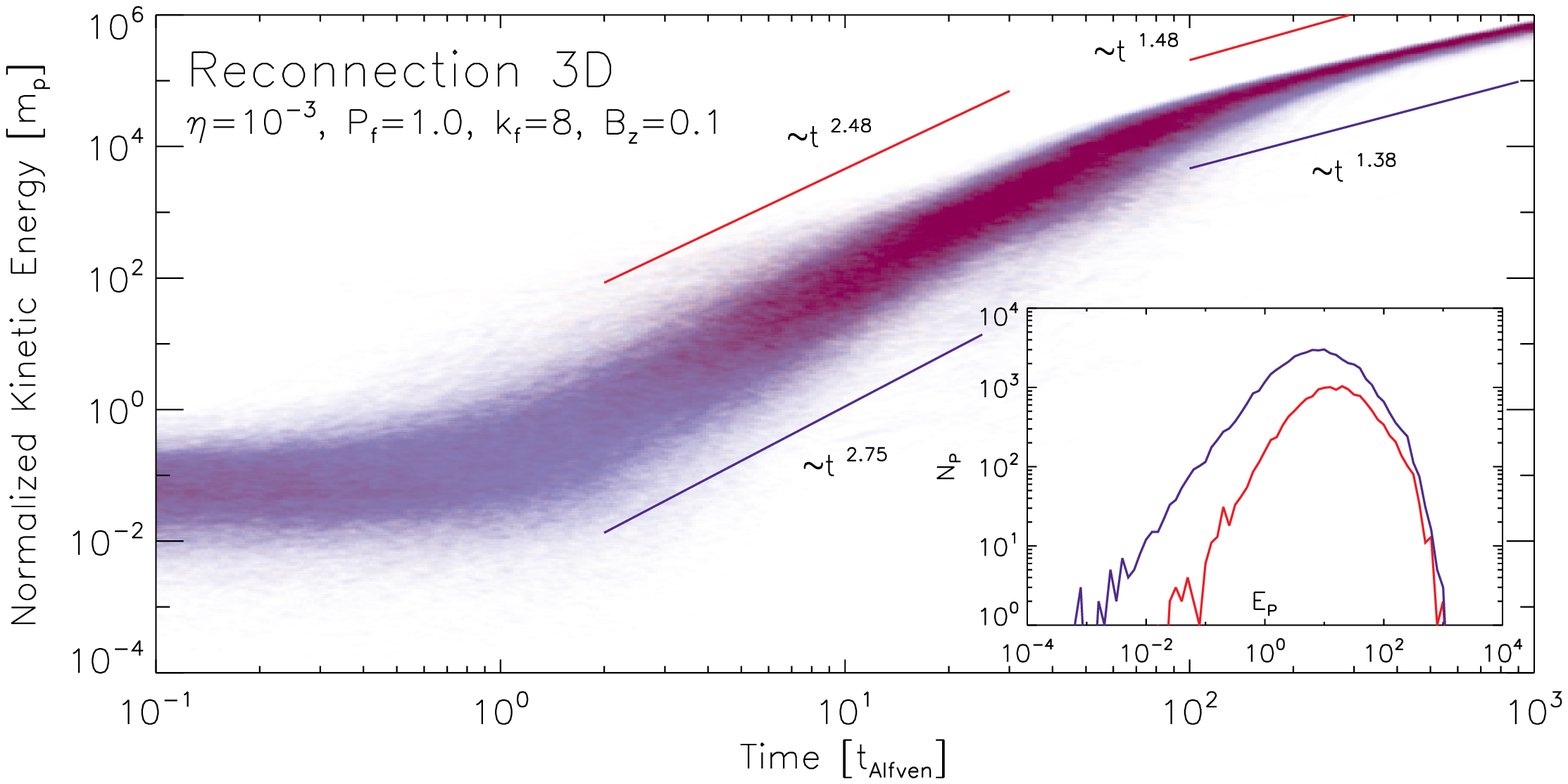}
\includegraphics[width=\columnwidth]{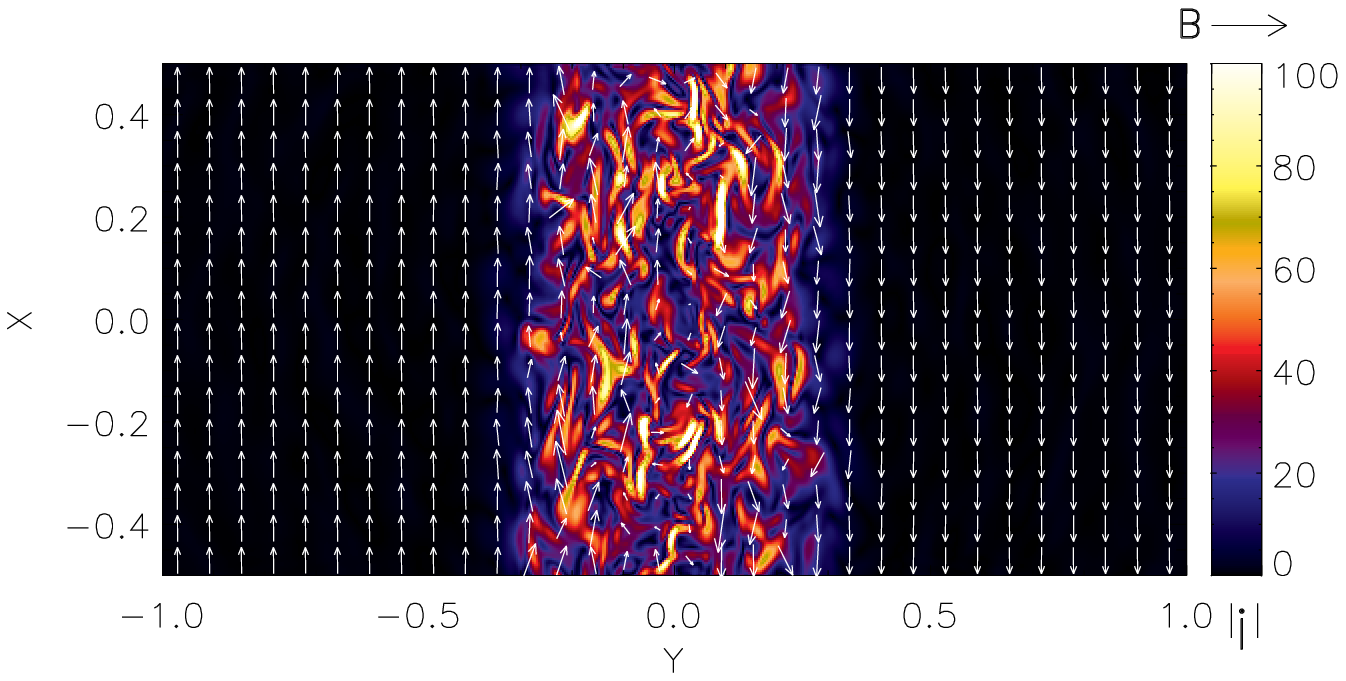} \\
\includegraphics[width=\columnwidth]{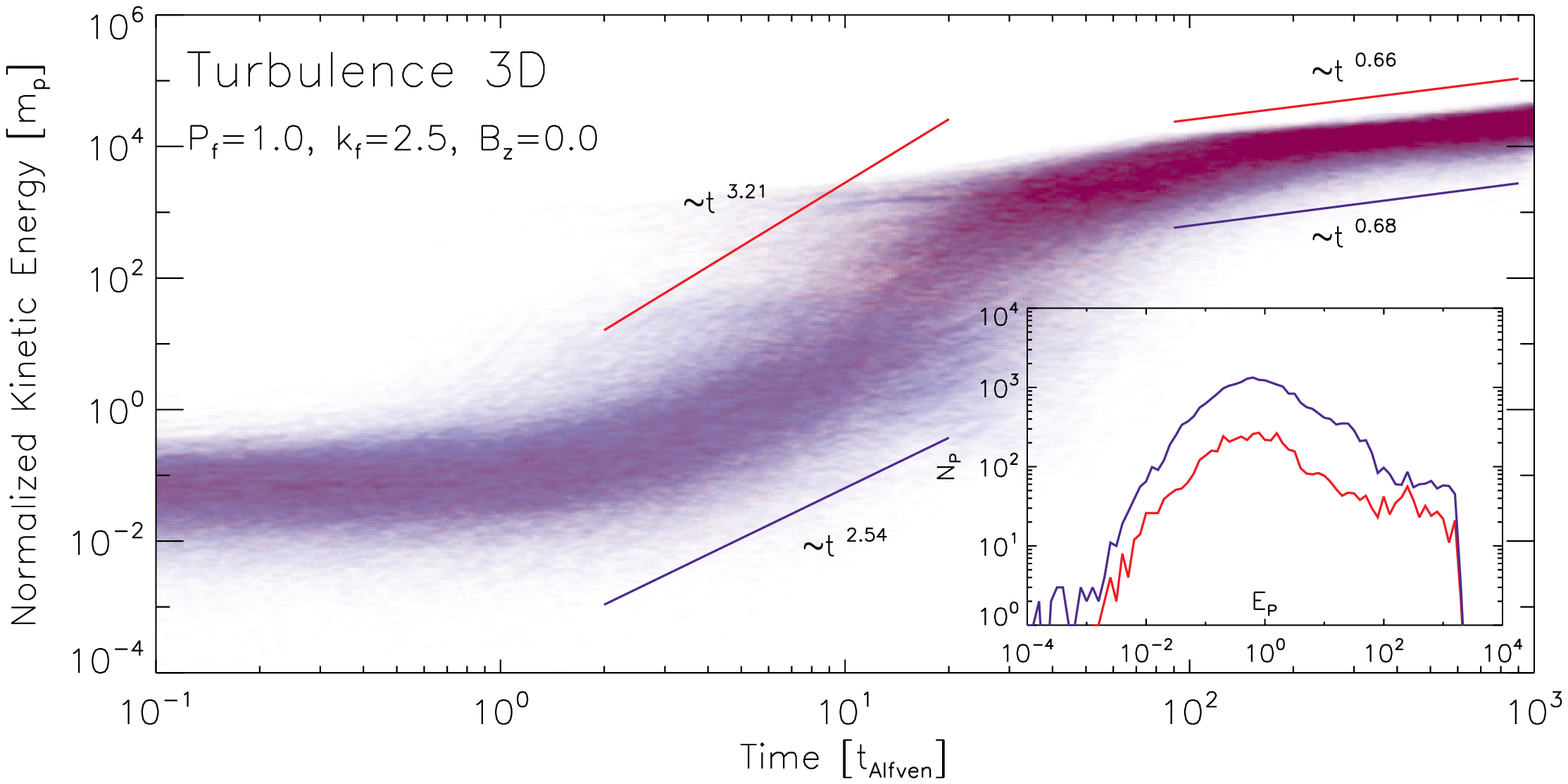}
\includegraphics[width=\columnwidth]{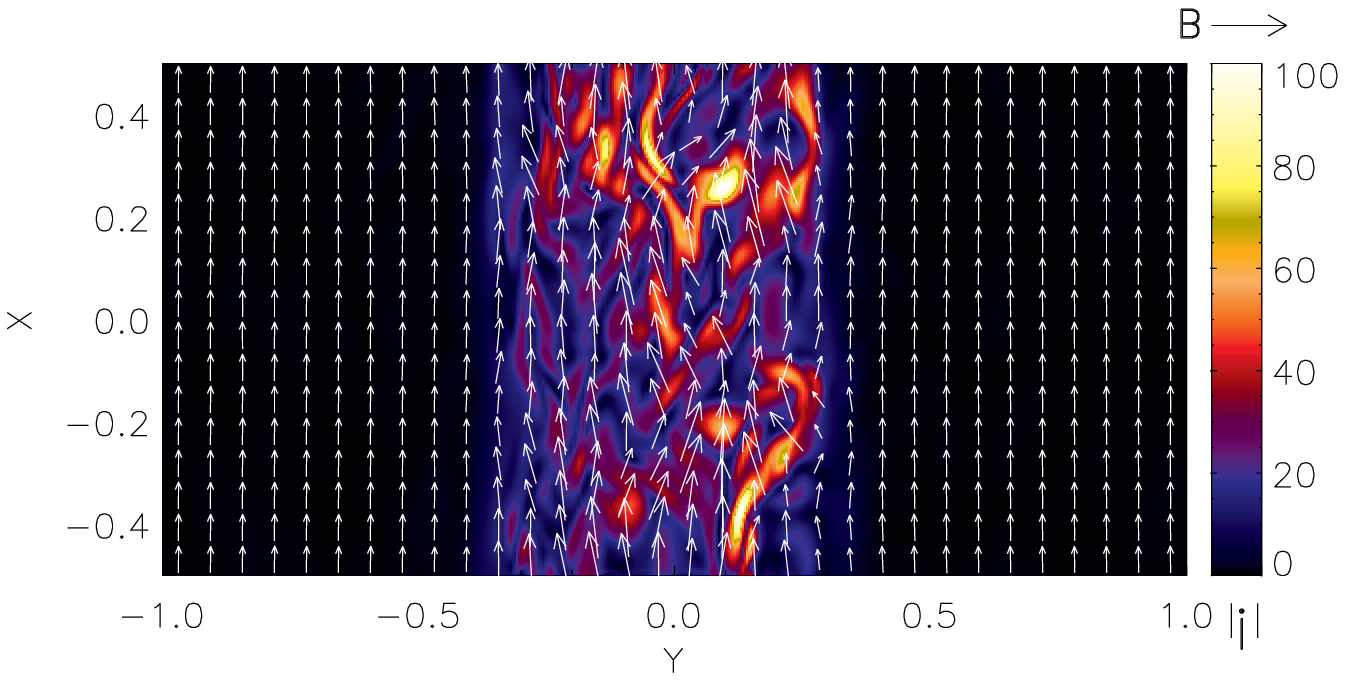}
\caption{{\em Left column:} Particle kinetic energy distributions for 10,000
protons injected in the Sweet-Parker reconnection (top), fast magnetic
reconnection (middle), and purely turbulent (bottom) domains.  The colors
indicate which velocity component is accelerated (red or blue for parallel or
perpendicular, respectively).  The energy is normalized by the rest proton mass.
Subplots show the particle energy distributions at $t=5.0$.  {\em Right column:}
The exemplary XY cuts through the domain at $Z=0$ of the absolute value of
current density $|\vec{J}|$ overlapped with the magnetic vectors for the
Sweet-Parker reconnection (top), fast reconnection (middle), and purely
turbulent domains (bottom).  Models with $B_{0z} = 0.1$, $\eta=10^{-3}$, and the
resolution 256x512x256 for reconnection and $B_{0z} = 0.2$ and the resolution
128x256x128 for the turbulent cases are shown. \label{fig:energy_evolution}}
\end{figure*}

In the Sweet-Parker model \citep{sweet58,parker57} the reconnection speed is
given by $V_{rec}\approx V_A S^{-1/2}$, where $S = L V_A / \eta$ is the
Lundquist number.  Because of the typical huge astrophysical sizes ($L$), $S$ is
also huge for Ohmic diffusivity values (e.g., for the ISM, $S \sim10^{16}$).
The Sweet-Parker reconnection is very slow, unless we use an artificially large
diffusivity.  In the model shown in the top panel of
Figure~\ref{fig:energy_evolution} we employed a diffusivity coefficient $\eta =
10^{-3}$ expressed in code units.  This value, due to the numerical diffusivity,
is several orders of magnitude larger than typical Ohmic diffusivity in
astrophysical environments, and makes the Sweet-Parker reconnection in the
simulation efficient.  The time evolution of the energy distribution for the
accelerating particles is shown for this model in the top left panel of
Figure~\ref{fig:energy_evolution}.  Initially, the perpendicular acceleration
dominates, because the volume in which we inject particles is much larger than
the current sheet region (shown in the right panel of the same figure).  The
perpendicular acceleration, due to a drift of the magnetic flux, starts before
the particles reach the reconnection region \cite{kowal11}.  The distribution of
particles does not change significantly until $t=1.0$.  Then, a rapid increase
in energy by roughly four orders of magnitude appears for a fraction of
particles.  We observe a big gap between the energy levels before and after
these acceleration events, which is also evident in the particle energy spectrum
depicted in the embedded subplot of the same diagram.  The events are spread in
time because particles gain substantial energy at different moments when
crossing the current sheet.  The energy growth during this stage is exponential.
This is a first-order Fermi acceleration process, as predicted in
\cite{degouveia05} and tested already in \cite{kowal11}.

Around $t=10$, we see a transition from the exponential to power-law particle
energy dependence with estimated index $\alpha \sim 1.1$ (see the upper plot of
Figure~\ref{fig:energy_evolution}).  A similar transition was also observed in
\cite{kowal11}.  In the plot we show the kinetic energy normalized by the proton
rest mass, what is equivalent to $(\gamma - 1)$.  The exponential acceleration
stops right after the energy value $10^4$ is reached, because the gyroradii
exceed the thickness of the acceleration region.  From this moment on the
particles cannot be confined within this thickness and the first order Fermi
process ceases.  Further energy increase is due to a much slower drift
acceleration (of the perpendicular component only) caused by the large scale
magnetic fields gradients.  The presence of a guide field allows the particles
to accelerate in the parallel direction as well.

Although the Sweet-Parker model with an artificially enhanced resistivity
results in a predominantly first-order Fermi acceleration, only a small fraction
of the injected particles is trapped and efficiently accelerated in the current
sheet (see the energy spectrum of the accelerated particles in the bottom right
of Figure~\ref{fig:energy_evolution}), because the acceleration zone is very
thin.

\paragraph{Acceleration in Reconnection with Turbulence.}
\label{sec:stochastic-reconnection}

Lazarian \& Vishniac \cite{lazarian99} proposed a model for fast reconnection
that does not depend on the magnetic diffusivity (see also \cite{eyink11}).  The
model appeals to the ubiquitous astrophysical turbulence as a universal trigger
of fast reconnection.  The predictions of this model have been successfully
tested in numerical simulations \cite{kowal09,kowal11b} which confirmed that the
reconnection speed is of the order of the Alfv\'en speed.  An important
consequence of the fast reconnection by turbulent magnetic fields is the
formation of a thick volume filled with small scale magnetic fluctuations.  In
order to test the acceleration of particles within such a domain, we introduced
turbulence within a current sheet with a Sweet-Parker configuration (as
described in the previous paragraph), then we followed the trajectories of
10,000 protons injected in this domain.

The middle left panel of Figure~\ref{fig:energy_evolution} shows the evolution
of the kinetic energy of the particles.  After injection, a large fraction of
test particles accelerates and the particle energy growth occurs earlier than in
the Sweet-Parker case (see also the energy spectrum at $t=5$ in the detail at
the bottom right of the same diagram).  This is explained by a combination of
two effects: the presence of a large number of converging small scale current
sheets and the broadening of the acceleration region due to the turbulence.
Here, we do not observe a gap seen in the Sweet-Parker reconnection, because
particles are continually accelerated by encounters with several small and
intermediate scale current sheets randomly distributed in the thick volume.  The
acceleration process is clearly still a first order Fermi process, as in the
Sweet-Parker case, but more efficient as it involves larger number of particles,
since the size of the acceleration zone and the number of scatterers have been
naturally increased by the presence of turbulence.  Moreover, the reconnection
speed, which in this case is independent of resistivity
\citep{lazarian99,kowal09} and determines the velocity at which the current
sheets scatter particles, has been naturally increased as well (i.e.
$V_{rec}\sim V_A$).

During this stage $\alpha$ is in the range $2.48-2.75$.  Then, like in the
laminar case, the protons accelerate at smaller rates after reaching energy
level $\sim 10^4$, because the thickness of the acceleration region (and of the
plasma scattering centers) becomes smaller than their Larmor radii.  This,
however, occurs later than in the previous case (around $t=100$).

\paragraph{Acceleration in Pure Turbulence.}
\label{sec:turbulence}

For comparison, in the bottom left panel of Figure~\ref{fig:energy_evolution} we
show the kinetic energy evolution of accelerated particles in a domain with
turbulence injected in an initially uniform large scale field.  In this domain
the acceleration is initially less efficient and a much smaller fraction of
particles is accelerated than in the reconnection case.  The increase in the
acceleration rate (with $\alpha = 2.54-3.21$) is in part due to constraints of
the computational domain.  We cannot inject turbulence at scales larger than the
size of the box.  As a consequence, an undesired converging flow arises along
the mean field due to large scale Alfv\'en waves, which enhances the
acceleration rate for particles with Larmor radii approaching the turbulent
injection scale.  After reaching an energy level of about $10^4$ proton mass,
this acceleration is significantly suppressed and $\alpha$ drops down to $\sim
0.67$.  In the model with fast reconnection the presence of the large scale
current sheet provides the converging flow.  This flow brings scattering centers
allowing a continuous growth of the particle energy until the saturation level.
In pure turbulence, the absence of a converging flow results in a random
particle scattering on approaching and receding small scale current sheets
(although at a smaller rate), so that the overall acceleration is a second-order
Fermi process.  This point still requires further studies, as reconnection
layers in pure turbulence can be responsible for first-order Fermi acceleration
of low energy particles.  As before, the rapid transition to smaller
acceleration rate occurs when the particle gyroradius reaches the size of the
turbulent domain and its irregularities (around $t=10$).  We note that we have
neglected here the time evolution of the MHD environment since this is much
longer than the particle time scales.  In fact, particles are accelerated by
magnetic fluctuations in the turbulent field and interact resonantly with larger
and larger structures as their energy increases due to the scatterings.  In a
steady state turbulent environment, as considered here, particles will see on
average the same sort of fluctuation distribution, so that after several
Alfv\'en times, one should expect no significant changes in the particle
spectrum due to the evolution of the large scale MHD environment.  Nonetheless,
this evolution may be important when considering more realistic non-steady
environments and when calculating real spectra and loss effects
\cite[e.g.][]{lehe09}.

\paragraph{Conclusions.}

In this letter we investigated particle acceleration in 3D MHD domains of
magnetic reconnection.  We found that the presence of turbulence significantly
increases the acceleration rate in a first-order Fermi process as predicted in
\cite{degouveia05}.  The particles trapped within the current sheet suffer
several head-on scatterings with the contracting magnetic fluctuations in the
thick volume.  In the Sweet-Parker model, where the reconnection speed was
artificially large due to numerical magnetic diffusivity, the acceleration rate
is slightly smaller because of a thinner current sheet.  This is still a
first-order Fermi process, however.  For comparison, we have also investigated
the acceleration in a pure 3D turbulence, where the particles with gyroradii
smaller than the injection scale accelerate through a second-order Fermi
process.

In summary, we have shown that the acceleration within reconnection sites,
especially in the presence of turbulence, can be extremely efficient.  This
could be a powerful mechanism not only in the solar corona and wind or the Earth
magnetotail, but also around black holes/accretion disks and jet launching
regions of AGNs and GRBs.  These results also call for further extensive work on
the CR acceleration in magnetic reconnection sites with the inclusion of
relevant loss mechanisms of CRs in order to assess the importance of this
acceleration mechanism in comparison to other processes (e.g., diffusive shock
acceleration) and to reproduce the observed light curves of the sources.

\paragraph{Acknowledgements.}

GK and EMGDP acknowledge support from FAPESP grants 2006/50654-3 and
2009/50053-8, and EMGDP from CNPq grant 300083/94-7.  AL acknowledges support
from CMSO, NSF AST-08-08118 and NASA A0000090101 grants, a Humboldt Award (Univ.
of Cologne and Univ. of Bochum), and from the International Institute of Physics
(Brazil).  Part of the computations were performed in GALERA supercomputer (ACK
TASK, Gda\'nsk, Poland) and using TACC resources (Teragrid AST080005N).



\begin{thebibliography}{}
\bibitem[de Gouveia Dal Pino \& Lazarian(2000)]{degouveia00}
 de Gouveia Dal Pino, E.~M. \& Lazarian, A.\ 2000, \apjl, 536, L31
\bibitem[de Gouveia Dal Pino \& Lazarian(2001)]{degouveia01}
 de Gouveia Dal Pino, E.~M. \& Lazarian, A.\ 2001, \apj, 560, 358
\bibitem[de Gouveia Dal Pino \& Lazarian(2005)]{degouveia05}
 de Gouveia Dal Pino, E.~M. \& Lazarian, A.\ 2005, \aap, 441, 845
\bibitem[de Gouveia Dal Pino et al.(2010a)de Gouveia Dal Pino, Piovezan \& Kadowaki]{degouveia10a}
 de Gouveia Dal Pino, E.~M., Piovezan, P.~P. \& Kadowaki, L.~H.~S.\ 2010a, \aap, 518, A5
\bibitem[de Gouveia Dal Pino et al.(2010b)de Gouveia Dal Pino,  Kowal, Kadowaki, Piovezan \& Lazarian]{degouveia10b}
 de Gouveia Dal Pino, E.~M.,  Kowal, G., Kadowaki, L. H. S., Piovezan, P., \& Lazarian, A.\ 2010b, \ijmpd, 19, 729
\bibitem[del Valle et al.(2011)del Valle, Romero, Luque-Escamilla, Mart{\'{\i}} \& Ram{\'o}n S{\'a}nchez-Sutil]{delvalle11}
 del Valle, M.~V., Romero, G.~E., Luque-Escamilla, P.~L., Mart{\'{\i}}, J., \& Ram{\'o}n S{\'a}nchez-Sutil, J.\ 2011, \apj, 738, 115
\bibitem[Drake et al.(2006)Drake, Swisdak, Schoeffler, Rogers \& Kobayashi]{drake06}
 Drake, J.~F., Swisdak, M., Schoeffler, K.~M., Rogers, B.~N., \& Kobayashi, S.\ 2006, \grl, 33, 13105
\bibitem[Drake et al.(2009)Drake, Cassak, Shay, Swisdak \& Quataert]{drake09}
 Drake, J.~F., Cassak, P.~A., Shay, M.~A., Swisdak, M., \& Quataert, E.\ 2009, \apjl, 700, L16
\bibitem[Drake et al.(2010)Drake, Opher, Swisdak \& Chamoun]{drake10}
 Drake, J.~F., Opher, M., Swisdak, M., \& Chamoun, J.~N.\ 2010, \apj, 709, 963
\bibitem[Eyink et al.(2011)]{eyink11}
 Eyink, G.~L., Lazarian, A., \& Vishniac, E.~T.\ 2011, \apj, 473, 51
\bibitem[Giannios(2010)]{giannios10}
 Giannios, D.\ 2010, \mnras, 408, L46
\bibitem[Gordovskyy et al.(2010)Gordovskyy, Browning \& Vekstein]{gordovskyy10}
 Gordovskyy, M., Browning, P.~K., \& Vekstein, G.~E.\ 2010, \apj, 720, 1603
\bibitem[Gordovskyy \& Browning(2011)]{gordovskyy11}
 Gordovskyy, M. \& Browning, P.~K.\ 2011, \apj, 729, 101
\bibitem[Kotera \& Olinto(2011)]{kotera11}
 Kotera, K. \& Olinto, A.~V.\ 2011, \araa, 49, 119
\bibitem[Kowal et al.(2011)Kowal, de Gouveia Dal Pino \& Lazarian]{kowal11}
 Kowal, G., de Gouveia Dal Pino, E.~M., \& Lazarian, A.\ 2011, \apj, 735, 102
\bibitem[Kowal et al.(2009)]{kowal09}
 Kowal, G., Lazarian, A., Vishniac, E.~T., \& Otmianowska-Mazur, K.\ 2009, \apj, 700, 63
\bibitem[Kowal et al.(2011)Kowal, Lazarian, Vishniac \& Otmianowska-Mazur]{kowal11b}
 Kowal, G., Lazarian, A., Vishniac, E.~T., \& Otmianowska-Mazur, K.\ 2011, \npg, in press
\bibitem[La Rosa et al.(2006)]{larosa06}
 La Rosa, T.~N., Shore, S.~N., Joseph, T., Lazio, W., \& Kassim, N.~E.\ 2006, \jpcs, 54, 10
\bibitem[Lazarian \& Vishniac(1999)]{lazarian99}
 Lazarian, A. \& Vishniac, E.~T.\ 1999, \apj, 517, 700
\bibitem[Lazarian \& Opher(2009)]{lazarian09}
 Lazarian, A. \& Opher, M.\ 2009, \apj, 703, 8
\bibitem[Lazarian \& Desiati(2010)]{lazarian10}
 Lazarian, A. \& Desiati, P.\ 2010, \apj, 722, 188
\bibitem[Lehe et al.(2009)]{lehe09}
 Lehe, R., Parrish, I.~J., \& Quataert, E.\ 2009, \apj, 707, 404
\bibitem[Lekien \& Marsden(2005)]{lekien05}
 Lekien, F., \& Marsden, J.\ 2005, \ijnme, 63, 455
\bibitem[Litvinenko(1996)]{litvinenko96}
 Litvinenko, Y.~E.\ 1996, \apj, 462, 997
\bibitem[Londrillo \& Del Zanna(2000)]{londrillo00}
 Londrillo, P., \& Del Zanna, L. 2000, \apj, 530, 508
\bibitem[Lyubarsky \& Liverts(2008)]{lyubarsky08}
 Lyubarsky, Y. \& Liverts, M.\ 2008, \apj, 682, 1436
\bibitem[Melrose(2009)]{melrose09}
 Melrose, D.~B.\ 2009, (arXiv:astro-ph.SR/0902.1803)
\bibitem[Mignone(2007)]{mignone07}
 Mignone, A.\ 2007, \jcph, 225, 1427
\bibitem[Ostrowski(2002)]{ostrowski02}
 Ostrowski, M.\ 2002, \app, 18, 229
\bibitem[Parker(1957)]{parker57}
 Parker, E.~N. 1957, \jgr, 62, 509
\bibitem[Press et al.(1992)Press, Teukolsky, Vetterling \& Flannery]{press92}
 Press, W.~H., Teukolsky, S.~A., Vetterling, W.~T., \& Flannery, B.~P., 1992, {\em Numerical recipes in C (2nd ed.): the art of scientific computing}, Cambridge University Press, New York, NY, USA
\bibitem[Sironi \& Spitkovsky(2009)]{sironi09}
 Sironi, L. \& Spitkovsky, A.\ 2009, \apj, 698, 1523
\bibitem[Sol et. al(2011)]{sol11}
 Sol, E. et al.\ 2011, \app, in press
\bibitem[Sweet(1958)]{sweet58}
 Sweet, P.~A. 1958, Conf. Proc. IAU Symposium 6, {\em Electromagnetic Phenomena in Cosmical Physics}, ed. B. Lehnert, (Cambridge, UK:Cambridge University Press), 123
\bibitem[Uzdensky(2011)]{uzdensky11a}
 Uzdensky, D.~A.\ 2011, \ssr, 101
\bibitem[Uzdensky \& McKinney(2011)]{uzdensky11b}
 Uzdensky, D.~A. \& McKinney, J.~C.\ 2011, \pop, 18, 042105
\bibitem[Zhang \& Yan(2011)]{zhang11}
 Zhang, B. \& Yan, H.\ 2011, \apj, 726, 90
\bibitem[Zharkova et al.(2011)]{zharkova11}
 Zharkova, V.~V., Arzner, K., Benz, A.~O., et al.\ 2011, \ssr, 159, 357
\end{thebibliography}
\end{document}